\documentclass[]{aa}
\usepackage[varg]{txfonts}
\usepackage{natbib}
\usepackage{ae,aecompl}
\usepackage{graphicx}   
\usepackage{amsmath}    
\usepackage{amssymb}    
\usepackage{color,verbatim,url}
\usepackage{tabularx}
\usepackage{xcolor}
\usepackage{booktabs}
\usepackage{pdflscape}
\usepackage[colorlinks=True,allcolors={blue}]{hyperref}

\usepackage{breqn}

\definecolor{pink}{rgb}{0.858, 0.188, 0.478}
\definecolor{purple}{RGB}{76, 0,153}

\newcommand{\bbf}[1]{{#1}}

\begin{document}
\defcitealias{hildebrandt/etal:2020}{H20}
\defcitealias{wright/etal:2020}{W20}

\title{KiDS+VIKING-450: Improved cosmological parameter constraints from redshift calibration with self-organising maps}
\titlerunning{KV450: SOM Cosmology}

\author{ Angus H. Wright\inst{1} 
\and Hendrik Hildebrandt\inst{1}
\and Jan Luca van den Busch\inst{1}
\and Catherine Heymans\inst{1,2}
\and \\
Benjamin Joachimi\inst{3}
\and Arun Kannawadi\inst{4}
\and Konrad Kuijken\inst{5}
}
\authorrunning{Wright et. al.}
\institute{
Ruhr-Universit\"at Bochum, Astronomisches Institut, German Centre for Cosmological Lensing (GCCL), \\
Universit\"atsstr. 150, 44801 Bochum, Germany. \email{awright@astro.rub.de} \and 
Institute for Astronomy, University of Edinburgh, Royal Observatory, Blackford Hill, Edinburgh EH9 3HJ, UK. \and 
Department of Physics and Astronomy, University College London, Gower Street, London WC1E 6BT, UK \and 
Department of Astrophysical Sciences, Peyton Hall, Princeton University, Princeton, NJ 08544, USA \and 
Leiden Observatory, Leiden University, Niels Bohrweg 2, 2333 CA Leiden, the Netherlands 
}

\date{Released 12/12/2121}


\graphicspath{{./figures/}}

%
%

\abstract{ 
We present updated cosmological constraints for the KiDS+VIKING-450 cosmic shear data set (KV450), estimated using
redshift distributions and photometric samples defined using self-organising maps (SOMs). Our fiducial analysis finds
marginal posterior constraints of $S_8\equiv\sigma_8\sqrt{\Omega_{\rm m}/0.3}=0.716^{+0.043}_{-0.038}$; smaller than,
but \bbf{otherwise} consistent with, previous work using this data set ($|\Delta S_8| = 0.023$). We analyse additional samples and
redshift distributions constructed in three ways: excluding certain spectroscopic surveys during redshift calibration,
excluding lower-confidence spectroscopic redshifts in redshift calibration, and considering only photometric sources
which are jointly calibrated by at least three spectroscopic surveys. In all cases, the method utilised here proves
robust: we find a maximal deviation from our fiducial analysis of \bbf{$|\Delta S_8| \leq 0.011$} for all samples defined and
analysed using our SOM. \bbf{To demonstrate the reduction in systematic biases found within our analysis, we highlight our results 
when performing redshift calibration without the DEEP2 spectroscopic data set. In this case we find marginal posterior
constraints of $S_8=0.707_{-0.042}^{+0.046}$; a} difference with respect to the fiducial \bbf{that} is both
significantly smaller than, and in the opposite direction to, the equivalent shift from previous work. These
results suggest that our improved cosmological parameter estimates are insensitive to pathological misrepresentation of
photometric sources by the spectroscopy used for direct redshift calibration, and therefore that this systematic effect
cannot be responsible for the observed difference between $S_8$ estimates made with KV450 and Planck CMB probes. 
}


\maketitle

\section{Introduction}
\label{sec: intro}
Estimation of cosmological parameters using tomographic cosmic shear requires accurate calibration of source redshift
distributions. For Stage III cosmic shear surveys such as the Kilo Degree Survey \citep[KiDS;][]{kuijken/etal:2019},
the Dark Energy Survey \citep[DES;][]{flaugher/etal:2015}, and the Hyper-Suprime Camera Wide-Survey
\citep[HSC;][]{aihara/etal:2018}, coherent biases on the order of $\Delta \langle z\rangle = \langle z \rangle_{\rm
est}-\langle z\rangle_{\rm true} \sim 0.04$ are enough to cause significant shifts in estimated cosmological parameter
estimates \citep[see, e.g,][]{ hildebrandt/etal:2017}. 
Systematic shifts of this nature are important given the observed (currently mild) tension between cosmological 
parameters estimated using KiDS weak lensing and cosmic microwave background (CMB) studies
\citep{planck/cosmo:2018}. 
For this reason, considerable effort has been invested in developing, testing, and optimising redshift calibration
methodologies for cosmic shear. 
These  methods can typically be grouped into three categories: those which
utilise cross-correlation \citep[see, e.g,][]{schneider/etal:2006,newman:2008,mcquinn/etal:2013,morrison/etal:2017}, 
stacking of individual redshift probability distributions \citep[see,
e.g,][]{hildebrandt/etal:2012,hoyle/etal:2018,tanaka/etal:2018}, or direct calibration using spectroscopic 
redshift training samples \citep[see, e.g,][]{lima/etal:2008,hildebrandt/etal:2017,hildebrandt/etal:2020,
buchs/etal:2019,wright/etal:2020}. \bbf{Recently, though, steps have been taken towards constructing a fourth, hybrid
category which leverages both cross-correlation and direct calibration
\citep{rau/etal:2019,sanchez/etal:2018,alarcon/etal:2019}.} 

The methodological differences, and implicit assumptions, between these estimation/calibration methods mean that they
are each susceptible to subtly different biases and systematic effects. For direct calibration methods, the completeness
and pre-selection of the spectroscopic training sample has been of particular concern \citep[see,
e.g,][]{gruenbrimioulle:2017,hartley/etal:2020}.  In \cite{wright/etal:2020} we developed an updated implementation of
the direct calibration procedure utilising self-organising maps \citep[SOMs;][]{kohonen:1982}, which we found to be less
susceptible to bias than previous implementations. We achieved this by the direct flagging and removal of photometric
sources which are not directly associated with a spectroscopic calibrator, thereby constructing a sample of fully
represented photometric sources and an associated redshift distribution: the `gold' sample. 

In this letter we apply the methodology of \cite{wright/etal:2020} to the KiDS+VIKING-450 data set of
\cite{wright/etal:2019}, and perform a tomographic cosmic shear analysis akin to that of \cite{hildebrandt/etal:2020}.
The data set used is described in section \ref{sec: data}, as is the definition of our various photometric and
spectroscopic analysis samples.  Our results are presented in section \ref{sec: results}, and we summarise the results
presented in this letter in section \ref{sec: summary}.

\section{Data set and Analysis Methodology}
\label{sec: data}
The KiDS+VIKING-450 data set (hereafter KV450) is presented in \cite{wright/etal:2019}, and \citet[][hereafter
\citetalias{hildebrandt/etal:2020}]{hildebrandt/etal:2020}. We utilise the cosmic shear data products from \citetalias{hildebrandt/etal:2020} with {\tt
lensfit} shape measurements \citep{miller/etal:2007,miller/etal:2013}, spectroscopic training samples 
\citep{vanzella/etal:2008,lilly/etal:2009,popesso/etal:2009,balestra/etal:2010,lefevre/etal:2013,
newman/etal:2013,kafle/etal:2018}, and BPZ photometric redshifts from \cite{benitez:2000}, as well as the core of the 
\citetalias{hildebrandt/etal:2020} 
parameter inference pipeline; we update only the redshift distributions using the new direct redshift calibration
methodology of \citet[][hereafter \citetalias{wright/etal:2020}]{wright/etal:2020}.  Our code is released as a stand-alone
analysis package\footnote{\url{https://www.github.com/AngusWright/CosmoPipe}}, with a wrapper
pipeline\footnote{\url{https://www.github.com/AngusWright/CosmoWrapper}} to perform the analyses presented in this work.
We provide the details of these cosmological analysis pipelines in Appendix \ref{app: pipelines}. 

In this analysis we utilise a range of differently compiled spectroscopic data sets to construct redshift
distributions and photometric source `gold samples' for cosmic shear analysis. A full description of the methods used to
construct these redshift distributions and gold samples is presented in \citetalias{wright/etal:2020}. Briefly, we
utilise self-organising maps (SOMs), trained on the \bbf{colours and magnitudes of sources in the} various spectroscopic
data sets, to associate photometric galaxies to spectroscopic galaxies with known redshift. \bbf{The SOM associations
are based on spectroscopic and photometric source assignments to specific SOM cells (galaxies falling within the same
SOM cell are assigned to the same associations; see Appendix B of \citetalias{wright/etal:2020}) and therefore links photometric and
spectroscopic sources with self-similar colour and magnitude properties.} Using these
associations, we are able to re-weight the spectroscopic redshift distribution to approximate the (unknown) photometric
galaxy redshift distributions. \bbf{Simultaneously, this} allows us to flag and remove photometric data which are not associated to spectra
(and therefore which are not represented by the re-weighted redshift distributions). 

\citetalias{wright/etal:2020} demonstrate that their redshift calibration methodology is less susceptible to systematic
biases in redshift distribution reconstruction, when compared with previously incorporated methods used by KiDS
(\citealt{hildebrandt/etal:2017}, \citetalias{hildebrandt/etal:2020}). \bbf{We are able to further this analysis, also using the simulations of
\cite{vandenbusch/etal:2020}, and estimate biases introduced by calibrating redshift distributions with
different spectroscopic calibration samples. This allows us to extend our cosmological analysis to include more informative
priors on the redshift distribution bias parameters, rather than incorporating only the zero-mean priors used by
\citetalias{hildebrandt/etal:2020}. Details of these new non-zero-mean priors (`$\delta z\neq0$') 
are given in Appendix \ref{app: priors}, and we adopt
the use of these improved priors for the majority of our analyses, where possible. In cases where we are unable to
derive updated redshift priors (because of limitations in the simulations of \citealt{vandenbusch/etal:2020}), we revert
to using the broader, zero-mean priors (`$\delta z=0$') from \citetalias{hildebrandt/etal:2020}.}

Finally, the construction of our gold photometric source subsamples requires the simultaneous recalibration of both
multiplicative and additive shear measurement bias parameters. While we are able to perform the additive shear bias
measurement on-the-fly within our cosmology pipeline, computation of the multiplicative shear biases is more involved.
We therefore pre-compute the required multiplicative shear bias values, using the methodology and simulations of
\cite{kannawadi/etal:2019}, for each of our photometric gold samples. These bias parameters are also given in Appendix
\ref{app: priors}. 

\subsection{Analysis samples}
In this work, we perform cosmic shear parameter estimation using a number of different photometric gold samples, redshift
distributions, and priors. \bbf{Following the analysis of \citetalias{wright/etal:2020},} our fiducial analysis defines
the gold sample as being those photometric data which are associated with one or more sources within the full KV450
spectroscopic compilation, and whose spectroscopic-to-photometric associations satisfy \bbf{an empirically derived
quality requirement. This requirement filters out associations whose mean photometric redshift $\langle Z^p_{\rm B}
\rangle_i$ \citep[from template fitting with BPZ;][]{benitez:2000} catastrophically disagrees with the mean
spectroscopic redshift of the association $\langle z^s_{\rm spec} \rangle_i$.  Such a criteria was found in
\citetalias{wright/etal:2020} to correlate strongest (in their simulations) with true bias of
photometric-to-spectroscopic associations, outperforming (in particular) the measured and true spectroscopic redshift
dispersions per association (which were found, counter-intuitively, to be uncorrelated with bias in the association
mean redshift estimates). We therefore implement the quality requirement: }
\begin{equation}\label{eqn: qc}
|\langle z^{s}_{\rm spec}\rangle_i - \langle Z^{p}_{\rm B}\rangle_i| \leq 
{\rm max}\left[5\times{\rm nMAD}\left(\langle z^s_{\rm spec}\rangle - \langle Z^{s}_{\rm B}\rangle\right),0.4\right],
\end{equation}
for each of the $i \in [1,N]$ association sets, where  $z^s_{\rm spec}$ is the spectroscopic redshift of the
spectroscopic sources, $Z^s_{\rm B}$ is the photometric redshift of the spectroscopic sources, and $Z^p_{\rm B}$ is the
photometric redshift of the photometric sources. This requirement is the same as presented in \citetalias{wright/etal:2020}, except that we have imposed a
floor on the threshold which defines catastrophic failure; we take as our threshold the maximum of $0.4$ and five times
the $z^s_{\rm spec} - Z^s_{\rm B}$ dispersion (determined using the normalised median absolute deviation from median; 
nMAD\footnote{$\sigma_{\rm nMAD}=1.4826\times{\rm med}\left(|x-{\rm med}(x)|\right)$. The pre-factor ensures normal
consistency; that is ${\rm E}[{\rm nMAD}(x_1,..,x_n)]= \sigma $ for $X\sim N(\mu,\sigma^2)$ and large $n$.}).
Redshift distributions are then calculated per tomographic bin ($Z_{\rm B} \in {(0.1,0.3],(0.3,0.5],(0.5,0.7]
,(0.7,0.9],(0.9,1.2]}$), as are the photometric gold samples. 

In addition to our fiducial analysis, we explore three gold samples constructed from spectroscopic compilations
excluding the zCOSMOS, VVDS, and DEEP2 data sets, respectively. We implement these samples both to compare with similar
samples run by \citetalias{hildebrandt/etal:2020}, and to test the sensitivity of our results to pathologically
under-representative spectroscopy. Further, we construct one gold sample (`specquality4') using only spectra which have
the highest quality flags from their various surveys (referred to as ${\rm nQ}>=4$ spectra, which have $\geq99.5\%$
confidence), to test the sensitivity of our analysis to spectra with a slightly higher likelihood of catastrophic
failures. Finally, we construct a highly restrictive gold sample (`multispec3') which consists only of sources which
reside in associations containing spectroscopy from (at least) three different spectroscopic surveys.
This selection, coupled with our quality control requirement, essentially restricts our sample to sources whose
calibration redshift is supported by multiple spectroscopic surveys with different selection functions, systematic
effects, and catastrophic failure modes. This calibration sample is therefore expected to be very robust (albeit at some 
cost to statistical precision due to a significant reduction in photometric effective number density);  
mis-calibration of these data would require coordinated catastrophic failure of redshift assignment across multiple
spectroscopic campaigns using different instruments and redshifting methods.

\begin{figure*}
\centering
\includegraphics[scale=0.9]{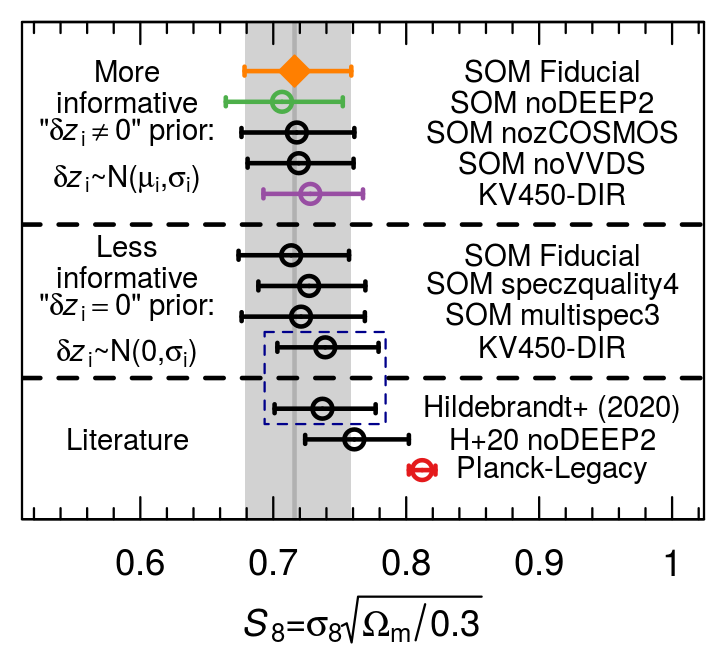}
\includegraphics[scale=0.9]{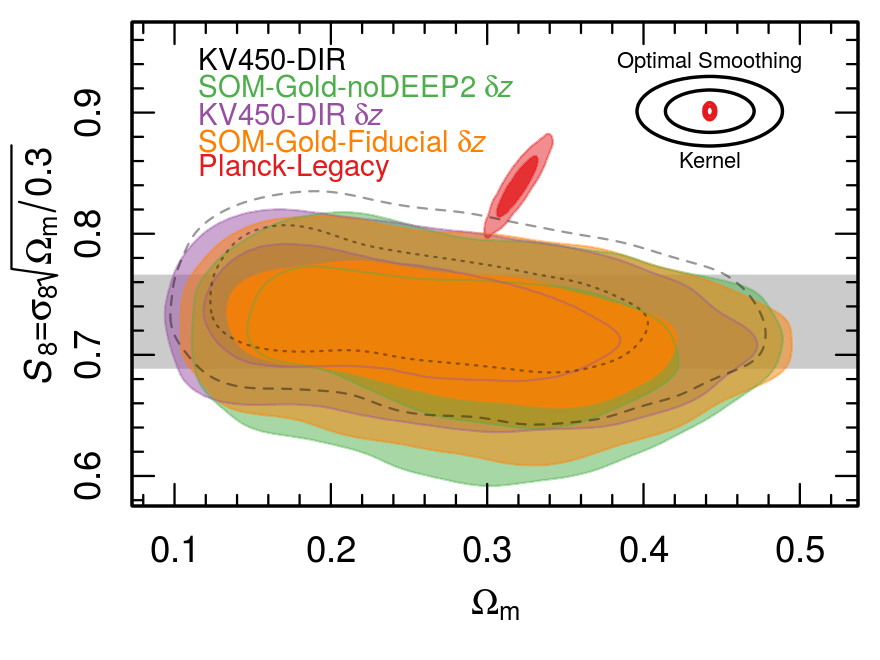}
\caption{Posterior constraints of $S_8$ (left) and $\Omega_{\rm m}$ vs. $S_8$ (right) for our various gold samples,
compared to the results from \protect \citetalias{hildebrandt/etal:2020} and Planck CMB. We show results for \bbf{our analyses
split into sections, determined by the form of their redshift distribution priors. Results computed 
using more informative}, non-zero mean, Gaussian redshift distribution bias priors (`$\delta z\neq0$', see Appendix \ref{app:
priors}) \bbf{for gold samples where these are able to be calculated}, and using the \bbf{zero mean Gaussian} bias priors from
\protect \citetalias{hildebrandt/etal:2020} otherwise (`$\delta z=0$'). \bbf{For our fiducial analysis we show results
with both priors, to allow direct comparison between our various results.} We annotate our contour figure (right) with
the two Gaussian smoothing kernels used in generating the contours (one for the cosmic shear contours, and one for the
CMB contours). We find that our new cosmology pipeline produces results consistent with the pipeline of \protect
\citetalias{hildebrandt/etal:2020} (left panel, blue dashed box). Our fiducial results (orange) suggest a slightly lower
$S_8$ than found in previous work: $S_8=0.716_{-0.038}^{+0.043}$. \bbf{However we find that constraints on $S_8$ are
extremely stable for all of our gold sample analyses here (compared to the fiducial: $|\Delta S_8|<0.2\sigma$),}
demonstrating that the results here are more robust to spectroscopic misrepresentation than previous works. \bbf{In
particular, unlike} \protect \citetalias{hildebrandt/etal:2020}, we find that even pathological misrepresentation at
high-redshift (`noDEEP2') is unable to shift our estimates of $S_8$ to larger values.  
}\label{fig: results}
\end{figure*}

\section{Results}\label{sec: results}
The results of our various gold sample cosmic shear measurements, quantified using the marginal posterior constraints of
the cosmic-shear summary parameter of interest $S_8=\sigma_8\sqrt{\Omega_{\rm m}/0.3}$, are shown in Figure \ref{fig:
results}.  Also shown are the results from \citetalias{hildebrandt/etal:2020} and Planck-Legacy
\citep{planck/cosmo:2018}, for comparison. The left panel is split into \bbf{three sections: analyses performed with
more informative non-zero mean redshift bias priors (`$\delta z \neq 0$', see Appendix \ref{app: priors}), 
estimated using the simulations of \citet{vandenbusch/etal:2020}; analyses performed with the broader, zero-mean redshift 
bias priors of \citetalias{hildebrandt/etal:2020} (`$\delta z = 0$'; used in cases where we cannot derive improved
priors or wish to compare directly to previous work); and external results taken directly from the
literature.} 

First, we verify our updated cosmology pipeline by performing an identical cosmological analysis to
\citetalias{hildebrandt/etal:2020}. As seen by the two results highlighted by the blue box in Figure \ref{fig: results},
we find that we recover essentially the same $S_8$ as they report: 
$S_8=0.739^{+0.040}_{-0.037}$ (labelled `KV450-DIR' in the figure, with $\delta z = 0$) compared to their
$S_8=0.737^{+0.040}_{-0.036}$ (`Hildebrandt+ (2020)'). We argue that the observed difference ($|\Delta S_8| \lesssim
0.003$) is simply a reflection of noise within our Markov-chain Monte-Carlo (MCMC).  For our fiducial gold sample
analysis, shown in orange in both
panels, we find a marginal constraint of $S_8=0.716_{-0.038}^{+0.043}$; smaller than, but \bbf{otherwise consistent
with, that which was found by \citetalias{hildebrandt/etal:2020}. Our} fiducial analysis is in better agreement with the results of
\citetalias{hildebrandt/etal:2020} when their data set and redshift distributions are  \bbf{re-analysed with more
informative redshift distribution bias priors (estimated from simulations, see Appendix \ref{app: priors}):} 
$S_8=0.727_{-0.036}^{+0.039}$ (`KV450-DIR' with $\delta z \neq 0$; purple). We
observe that our fiducial analysis has a slightly broader marginal $S_8$ constraint.  This is expected when
performing our gold selection: by decreasing the size of the photometric data set which is used for the analysis (which
we quantify using the change in the effective number density of cosmic shear source galaxies, $\Delta n_{\rm eff}=n^{\rm
gold}_{\rm eff}/n^{\rm all}_{\rm eff}\approx80\%$ for our fiducial sample; see Appendix \ref{app: neff}), we increase
the statistical noise on our marginal constraints. 

We explore the sensitivity of our analysis to the construction of our spectroscopic compilation, by performing our
analysis with gold samples constructed without spectra from zCOSMOS, VVDS, and DEEP2. When removing  zCOSMOS or VVDS,
we find that our marginal constraint on $S_8$ is unchanged within MCMC noise: $|\Delta S_8|
\lesssim 0.003$. In the cases of removing DEEP2 from the calibration sample, we find a shift in our marginal
constraint of $S_8=0.707^{+0.046}_{-0.042}$, equating to $|\Delta S_8| \lesssim 0.2\sigma$. We note
though, that (looking at the $\Omega_{\rm m}$ versus $S_8$ plane) we can see that the shift in $S_8$ without DEEP2 is
driven by an extension of the posterior to lower values, rather than a systematic biasing of the distribution overall. 
\bbf{This result is of further interest in the context of the full multi-dimensional posterior space, which we present
in Appendix \ref{app: marginals} for the interested reader.} 

We draw particular attention to the differences seen between our analysis without DEEP2 and the \bbf{equivalent} analysis performed
by \citetalias{hildebrandt/etal:2020}. 
When performing their noDEEP2 analysis \citetalias{hildebrandt/etal:2020} found a non-trivial increase in $S_8$
to $S_8=0.761^{+0.041}_{-0.037}$ (`H+20 noDEEP2'); a shift of $\Delta S_8\sim +0.6\sigma$. 
This difference is attributed, in \citetalias{hildebrandt/etal:2020}, to a bias in the reconstructed redshift
distributions used for this test: removing DEEP2 causes pathological misrepresentation of the high-redshift portion of
the spectroscopic colour-colour space, which subsequently causes the reconstructed redshift distributions to be
systematically biased low, thereby introducing a positive shift in $S_8$ for the otherwise unchanged photometric source
sample. \bbf{To this end, \citetalias{hildebrandt/etal:2020} argued that their noDEEP2 analysis was systematically
biased, and was a demonstration of the limitations of their redshift calibration procedure. In the SOM based gold
samples, however, we find no such biasing of our results when calibrating without spectra from DEEP2.} 
We attribute this difference to the gold selection
process: our redshift distributions without DEEP2 are similarly skewed low compared to the fiducial (see Appendix
\ref{app: neff}), however unlike \citetalias{hildebrandt/etal:2020} our gold selection simultaneously removes the
misrepresented photometric sources. Therefore, while our redshift distributions change significantly between the
fiducial and noDEEP2 analyses, both correctly describe the photometric data within their respective gold samples; both
are accurate and consistent.  We therefore see no biasing of the derived cosmological parameters, but rather just an
increase in marginal uncertainties due to the aforementioned decrease in statistical power due to the
$\sim20\%$ reduction in the effective number density of the photometric sample. 

In addition to our tests for the effect of pathological colour misrepresentation, we also test the influence of
spectra which may have an increased fraction of catastrophic failures. Recall that in our spectroscopic compilation for
KV450 we allow only high-confidence ($\geq95\%$) and/or `certain' ($\geq99.5\%$ confidence) spectra; however even
high-confidence spectra may have catastrophic failures. In \citetalias{wright/etal:2020} we demonstrated using
simulations that expected fractions of catastrophic spectroscopic failures were unlikely to bias calibration of redshift
distributions in KV450. Nonetheless, here we explore the influence of the lower-confidence spectroscopy on our
conclusions. Our `speczquality4' gold sample is calibrated using only certain confidence redshifts. \bbf{The resulting
marginal constraints of $S_8$ differ from our fiducial results by $|\Delta S_8|=0.011$.
We therefore conclude that the presence of lower-confidence spectra in our calibration data set does not
introduce significant biases in our fiducial marginal constraints of $S_8$.}

Our speczquality4 result is of additional interest in the context of recent work presented by \cite{hartley/etal:2020}. 
For DES (i.e. using fewer photometric bands than used in KiDS), and implementing a redshift calibration methodology akin
to that of \citetalias{hildebrandt/etal:2020}, they find switching between direct calibration using high-confidence
($\geq95\%$) and certain ($\geq99.5\%$) spectroscopic samples results in a significant $\Delta \langle z \rangle 
\geq 0.06$ bias for their highest tomographic bin ($Z_{\rm B}\in\left(0.7,1.3\right]$). While these biases are not
directly applicable to our analysis, any similar systematic bias within our analysis would likely cause a significant
change in the estimated cosmological parameters. We find no such significant bias when switching between direct 
calibration using high-confidence and certain spectroscopic redshifts, suggesting that this bias is suppressed in our
data set. We hypothesise that this is driven by one, or a combination, of the following three effects. Firstly, that
our 9-band photometric space is more resilient to spectroscopic selection biases than the 4-band space considered in 
\cite{hartley/etal:2020}. Secondly, that our deeper and more diverse spectroscopic compilation reduces the
sensitivity of the recalibration procedure to strong (survey-specific) spectroscopic selection effects. Finally, that
the calibration method of \cite{wright/etal:2020} is more resilient to spectroscopic selection effects than the method
used in \cite{hartley/etal:2020}. We leave exploration of these three possibilities to future work. 

\bbf{A change in marginal $S_8$ constraints of $|\Delta S_8|=0.011$, with respect to the fiducial case, is nonetheless large 
in the context of our gold samples. As such,} we extend this test further by implementing more stringent requirements on spectroscopic agreement. Our
`multispec3' gold sample consists only of photometric sources which are calibrated by spectra originating from at least
3 different spectroscopic surveys within our compilation. As stated in Section \ref{sec: data}, this requirement places
a strong restriction on spectroscopic agreement when coupled with our quality control requirement (Equation \ref{eqn:
qc}). For our multispec3 gold sample we find again a result which is consistent with our fiducial analysis: \bbf{ 
$S_8=0.721^{+0.048}_{-0.045}$, corresponding to $|\Delta S_8| \lesssim 0.005$, only slightly larger than the MCMC noise 
threshold. This result demonstrates the stability of our gold sample results,} and provides a strong indication that the
marginal constraints on $S_8$ presented here are not biased by systematic effects nor catastrophic failures within the
spectroscopic calibration sample. 

\bbf{While we have focussed our discussion here on the marginal $S_8$ constraints, in Appendix \ref{app: marginals} we
provide additional marginal and two-dimensional posterior constraints for a subset of our parameter distributions. We explore other
conclusions which we can draw from our gold cosmological analyses, specifically around intrinsic alignments and the
posterior probability distributions of $\Omega_{\rm m}$ and $\sigma_8$. Briefly, our gold sample marginal constriants show a reduced 
preference for low values of $\Omega_{\rm m}$, causing a more consistent recovery of $\Omega_{\rm m}\approx0.3$. 
In all of our gold analyses the marginal constraints are good agreement ($|\Delta X| < 0.2\sigma$ for all parameters $X$), 
with the exception of the intrinsic alignment amplitude parameter $A_{\rm IA}$, which shows up to $|\Delta A_{\rm
IA}|\sim 1.0\sigma$ differences among analyses. Importantly, though, our gold sample $A_{\rm IA}$ constraints are all
consistent with $A_{\rm IA}=0$, unlike those from \citetalias{hildebrandt/etal:2020}, who found $A_{\rm IA}\approx1$.
This updated constraint is in better agreement with recent work on intrinsic alignments \citep{fortuna/etal:2020}, who
predict an intrinsic alignment amplitude for KiDS of $0\leq A_{\rm IA}\leq0.2$. Finally, we find that differences in
$S_8$ and $A_{\rm IA}$ are correlated between our gold sample analyses, and that stronger prior information on $A_{\rm
IA}$ would further improve the agreement between our respective gold sample analyses.}

\section{Summary}\label{sec: summary}
We present updated cosmological parameter constraints from the KiDS+VIKING-450 data set of \cite{wright/etal:2019},
estimated using updated redshift distributions following the methodology of \cite{wright/etal:2020}. For our fiducial
analysis we find a value of $S_8$ that is smaller than, but nonetheless fully
consistent with, the value reported in the previous KiDS+VIKING-450 cosmological analysis of \cite{hildebrandt/etal:2020}:
$S_8=0.716^{+0.043}_{-0.038}$ compared to $S_8=0.737^{+0.040}_{-0.036}$ ($|\Delta S_8|\leq0.6\sigma$). We note, however,
that when one analyses the data and redshift distributions of \cite{hildebrandt/etal:2020} using updated redshift
distribution bias parameters presented in \cite{wright/etal:2020}, their $S_8$ also shifts downward and is in better
agreement with our fiducial analysis: $S_8=0.727^{+0.039}_{-0.036}$, $|\Delta S_8|\leq0.3\sigma$.  We explore the
sensitivity of our results to systematic misrepresentation within the spectroscopic calibration data set by removing
multiple spectroscopic subsamples (DEEP2, VVDS, zCOSMOS), each of which uniquely calibrate distinct portions of the
colour-redshift space. We find that the results
presented here are robust to pathological misrepresentation, whereby even the removal of DEEP2 is unable to cause a
significant shift in $S_8$: $|\Delta S_8|\leq0.2\sigma$. \bbf{We
find that our results are consistent with the fiducial when performing the calibration using only certain (nQ$=4$,
$\geq99.5\%$ confidence) spectroscopic redshifts, and when performing an extremely conservative analysis considering only
photometric sources which are simultaneously calibrated by spectra from at least three different spectroscopic
surveys: $|\Delta S_8|\leq0.011$ and $|\Delta S_8|\leq0.005$ respectively.} Overall these
results indicate that, using the redshift calibration methodology of \cite{wright/etal:2020}, pathological
misrepresentation of photometric sources within the spectroscopic compilation is not able to produce significant changes
in marginal constraints of $S_8$, and therefore cannot reconcile the $\Delta S_8\approx 2.5\sigma$ differences observed
between cosmological parameters estimated using KiDS and Planck.

\begin{acknowledgements}
\bbf{We thank the anonymous referee for their comments and suggestions, which have undoubtedly improved the quality of this
manuscript.} 
We acknowledge support from the European Research Council under grant numbers 770935 (AWH, HH, JvdB) and 647112 (CH).
HH is also supported by a Heisenberg grant (Hi1495/5-1) of the Deutsche Forschungsgemeinschaft.  CH also acknowledges
support from the Max Planck Society and the Alexander von Humboldt Foundation in the framework of the Max
Planck-Humboldt Research Award endowed by the Federal Ministry of Education and Research. KK acknowledges support from
the Humboldt Foundation, and the hospitality of Imperial College London.  This work is based on observations made with
ESO Telescopes at the La Silla Paranal Observatory under programme IDs 100.A-0613, 102.A-0047, 179.A-2004, 177.A-3016,
177.A-3017, 177.A-3018, 298.A-5015.  The MICE simulations have been developed at the MareNostrum supercomputer (BSC-CNS)
thanks to grants AECT-2006-2-0011 through AECT-2015-1-0013. Data products have been stored at the Port d'Informació
Científica (PIC), and distributed through the CosmoHub webportal (cosmohub.pic.es).
\end{acknowledgements}

\bibpunct{(}{)}{;}{a}{}{,}
\bibliographystyle{aa}
\bibliography{library}

\clearpage
\appendix 

\section{Cosmology and wrapper pipelines}\label{app: pipelines}
With this letter we release a new implementation of the KiDS cosmological analysis pipeline utilised by
\citetalias{hildebrandt/etal:2020}, which has been
generalised for ease of use. The new pipeline, simply called {\tt CosmoPipe}, is available from
\url{https://www.github.com/AngusWright/CosmoPipe}. The package can be installed trivially with the provided master
installation script, and generates a clean working environment for each installation of the pipeline to avoid conflicts
with, for example, existing python installations. 

{\tt CosmoPipe} contains the same analysis steps performed by \citetalias{hildebrandt/etal:2020}. The pipeline utilises
{\tt treecorr} \citep{jarvis/etal:2004} for computation of cosmic shear correlation functions, and 
{\tt MontePython} \citep{audren/etal:2013} for Markov-chain
monte-carlo (MCMC) analyses. For clarity, we outline the seven primary steps of the pipeline here. 
\begin{enumerate}
\item Compute the 2D c-term for all survey patches and tomographic bins;
\item Compute the 1D c-term for all survey patches;
\item Compute 2pt shear correlation functions;
\item Construct the correlation function covariance matrix;
\item Prepare the data for input to {\tt MontePython} MCMC: reformat the correlation functions, reformat the covariance
matrix, prepare the montepython likelihood, reformat the Nz distributions, define the correlation function scale-cuts, 
and link any required treecorr files; 
\item Run the MCMC; 
\item Construct summary figures and statistics from the MCMC chains.  
\end{enumerate}
While this pipeline has been largely generalised, it is clear that some of these steps above are tailored for KiDS-like
cosmological analyses. For example, the {\tt CosmoPipe} is provided with a version of the public KiDS likelihood that has
been pre-formatted to fit seamlessly into the {\tt CosmoPipe}. The code will function equally well with an arbitrary
likelihood, albeit with some additional preparation required on the user-side.  

Should one wish to perform an analysis such as (or indeed identical to) that presented here, we also provide a wrapper
package which links together the cosmological analysis pipeline package and the redshift calibration package of
\citetalias{wright/etal:2020}. This wrapper package, available at \url{https://www.github.com/AngusWright/CosmoWrapper},
contains one main script, {\tt Wright2020b.sh}, which performs the entirety of the analysis presented here. This script
requires only that the user have the input photometric and spectroscopic calibration data sets supplied, and performs
(with one command) the full gambit of analysis required for this letter.  These steps include: 
\begin{enumerate}
\item redshift calibration;
\item gold sample selection; 
\item installation of {\tt CosmoPipe}; 
\item preparation of {\tt CosmoPipe} for the different gold sample runs; 
\item running {\tt CosmoPipe}; and
\item outputting of figures present in this paper. 
\end{enumerate}
Some additional input parameters to the {\tt CosmoPipe} are also encoded in the wrapper package, such as the various
redshift distribution and multiplicative shear bias priors given discussed in Appendix \ref{app: priors}.

\section{Gold sample priors}\label{app: priors}
\subsection{Mean redshift distribution biases}
\vspace{5pt}
\begin{table*}
\centering
\caption{Updated redshift distribution bias priors parameters used in different gold sample analyses. Priors are
Gaussian ($\mu \pm \sigma$). Parameters are determined from the simulations of \protect \cite{vandenbusch/etal:2020} as
described in \citetalias{wright/etal:2020}, except that we double the simulation bias standard deviations when constructing
our priors. For samples where we want to replicate previous analyses, we implement the prior from
\citetalias{hildebrandt/etal:2020} (`All $\delta z=0$'). }\label{tab: dzpriors}
\begin{tabular}{cc|r@{\hskip 0in}lr@{\hskip 0in}lr@{\hskip 0in}lr@{\hskip 0in}lr@{\hskip 0in}l}
\hline
\multicolumn{2}{c|}{Gold} & \multicolumn{10}{c}{Tomographic Redshift Bias Prior $\delta z$} \\
\multicolumn{2}{c|}{Sample}& \multicolumn{2}{c}{bin1} & \multicolumn{2}{c}{bin2} & \multicolumn{2}{c}{bin3} &
\multicolumn{2}{c}{bin4} & \multicolumn{2}{c}{bin5} \\
&            & \multicolumn{2}{c}{$Z_{\rm B}\in (0.1,0.3]$} & \multicolumn{2}{c}{$(0.3,0.5]$} &
            \multicolumn{2}{c}{$(0.5,0.7]$} & \multicolumn{2}{c}{$(0.7,0.9]$} & \multicolumn{2}{c}{$(0.9,1.2]$} \\
\hline
&Fiducial    &$ 0.000 $&$\pm 0.010 $&$ 0.002 $&$\pm 0.012 $&$ 0.013 $&$\pm 0.012 $&$ 0.011 $&$\pm 0.008 $&$-0.006 $&$\pm 0.010 $\\
&KV450-DIR   &$ 0.047 $&$\pm 0.010 $&$ 0.025 $&$\pm 0.008 $&$ 0.032 $&$\pm 0.010 $&$-0.004 $&$\pm 0.008 $&$-0.013 $&$\pm 0.008 $\\
$\delta z \neq 0$&NoDEEP2     &$-0.001 $&$\pm 0.010 $&$ 0.002 $&$\pm 0.012 $&$-0.002 $&$\pm 0.012 $&$-0.009 $&$\pm 0.010 $&$-0.015 $&$\pm 0.010 $\\
&noVVDS      &$ 0.001 $&$\pm 0.010 $&$ 0.001 $&$\pm 0.012 $&$ 0.024 $&$\pm 0.014 $&$ 0.014 $&$\pm 0.010 $&$-0.007 $&$\pm 0.012 $\\
&nozCOSMOS   &$ 0.005 $&$\pm 0.026 $&$ 0.005 $&$\pm 0.016 $&$ 0.032 $&$\pm 0.014 $&$ 0.030 $&$\pm 0.010 $&$ 0.002 $&$\pm 0.012 $\\
\hline
\multicolumn{2}{c|}{All $\delta z=0$}& $0.000 $&$\pm 0.039$ & $0.000 $&$\pm 0.023$ & $0.000 $&$\pm 0.026$ & $0.000 $&$\pm 0.012$ & $0.000 $&$\pm 0.011$ \\
\hline
\hline
\end{tabular}
\end{table*}
\bbf{Redshift distribution bias is a significant uncertainty in cosmic shear analyses, as coherent bias in redshift
distribution means is directly reflected as a bias in $S_8$. In \citetalias{hildebrandt/etal:2020}, redshift
distributions were assigned a zero-mean (`$\delta z=0$') Gaussian prior, whose width (per tomographic bin) was estimated via a spatial
bootstrap method (see Section 3.2 of \citetalias{hildebrandt/etal:2020}). These priors are applicable to the direct
calibration implemented in their work, but need not be reflective of the biases in our SOM based redshift calibration
procedure. 

Fortunately, \citetalias{wright/etal:2020} present estimates of the redshift calibration biases for our fiducial gold
sample. Briefly, these redshift distribution bias estimates are made
using simulations of the KiDS photometric and spectroscopic compilations from \cite{vandenbusch/etal:2020}. 
These simulations generate 100 independent realisations of
the three main spectroscopic lines-of-sight within KiDS, which are then used to calibrate a mock KiDS photometric
sample. At all steps in this process, samples are constructed to match the real KiDS photometric and spectroscopic data
in all possible manners: magnitude, colour, selection effects, weights, redshift distributions, etc. This simulation
therefore allows the estimation of mean redshift distribution biases, include uncertainties and biases introduced by
photometric noise, sample variance, spectroscopic selection effects, spectroscopic incompleteness, and Poisson sampling.
Tests performed in \citetalias{wright/etal:2020} suggest that the dominant sources of calibration uncertainty are
photometric noise and spectroscopic selection effects. Using the recovered mean redshift distribution bias estimates
(and uncertainties), we are able to construct new, more informative, non-zero-mean (`$\delta z\neq0$') redshift
distribution bias priors for our cosmic shear analysis. These priors are presented in Table \ref{tab: dzpriors}.

As each of the zCOSMOS, VVDS, and DEEP2 samples were simulated by \cite{vandenbusch/etal:2020}, we are able to also 
calibrate the redshift bias priors for the three gold samples which exclude these subsamples: our `nozCOSMOS',
`noVVDS', and `noDEEP2' gold samples. However, the simulations of \cite{vandenbusch/etal:2020} did not include estimates
of redshift quality, so we are unable to construct a more informative prior for the `speczquality4' gold sample.
Similarly, as they do not simulate every spectroscopic subsample within KiDS, construction of a new `multispec3' prior
is also not possible. In each of these cases, we are therefore required to default back to the `$\delta z=0$' priors of
\citetalias{hildebrandt/etal:2020}. 

To be
conservative, we opt to double the uncertainties on the bias found in the simulations when constructing our priors. We
opt to utilise these new priors, where possible, as they represent our current best-estimate
of the true redshift bias parameters inherent to the recalibration method and samples used here, despite the limitations
of the simulations used \citep{vandenbusch/etal:2020,wright/etal:2020}. We note, however, that the biases are typically
small, being of order $\delta z \lesssim 0.01$ for the majority of samples and bins, meaning that the
\citetalias{hildebrandt/etal:2020} $\delta z=0$ prior is nonetheless a reasonable approximation (and therefore unlikely
to be a source of bias in the cases where we are required to use that prior; i.e. speczquality4 and multispec3). }

\subsection{Multiplicative shear bias}\label{app: mbiaspriors}
As each of our gold selections produces a different subset of the full photometric sample, this requires a new
computation of the multiplicative and additive shear biases, shear correlation functions, covariances, etc. 
Each of these is incorporated into the pipeline processing, with the exception of the multiplicative shear bias 
estimation. In \citetalias{hildebrandt/etal:2020}, multiplicative shear
biases are computed using the methodology and simulations of \cite{kannawadi/etal:2019}. We invoke the same procedure,
thereby generating a bespoke set of multiplicative shear-bias parameters for each of our gold selections, albeit 
outside of our wrapper pipeline. These bias parameters are given in Table \ref{tab: mbias} for each of our gold samples.
We note that in all cases we have chosen to implement the same m-bias uncertainty as used in
\citetalias{hildebrandt/etal:2020}: $\Delta m = 0.02$ for all tomographic bins.

We recognise that this implementation of the m-bias estimation may be sub-optimal: the gold selections are strongly
colour-dependent, and our current simulation-set for estimating multiplicative bias is entirely mono-chromatic (using
only the $r$-band imaging and fluxes). However these simulations nonetheless represent the state-of-the-art within
KV450, and we leave exploration of how the m-biases change with multi-colour simulations for future studies.  These
m-bias parameters are required as input to our cosmology pipeline, and so are documented here. Overall, the gold sample
multiplicative biases are very similar, with only the multispec3 calibration differing from the fiducial by more than
$\Delta m \sim 0.002$. In all cases, the different m-bias values are well within the assumed multiplicative bias
uncertainty used here. 

\begin{table*}
\centering
\caption{Multiplicative shear bias parameters used for each of our gold sample analyses.} \label{tab: mbias}
\begin{tabular}{c|rrrrr}
Gold   & \multicolumn{5}{c}{multiplicative shear bias parameter} \\
Sample & \multicolumn{1}{c}{bin1} & \multicolumn{1}{c}{bin2} & \multicolumn{1}{c}{bin3} & \multicolumn{1}{c}{bin4} &
\multicolumn{1}{c}{bin5} \\
\hline
Fiducial     &$-0.0145 \pm 0.0200$&$-0.0176 \pm 0.0200$&$-0.0125 \pm 0.0200$&$ 0.0045 \pm 0.0200$&$ 0.0122 \pm 0.0200$\\
NoDEEP2      &$-0.0137 \pm 0.0200$&$-0.0162 \pm 0.0200$&$-0.0112 \pm 0.0200$&$ 0.0054 \pm 0.0200$&$ 0.0130 \pm 0.0200$\\
noVVDS       &$-0.0143 \pm 0.0200$&$-0.0172 \pm 0.0200$&$-0.0116 \pm 0.0200$&$ 0.0047 \pm 0.0200$&$ 0.0125 \pm 0.0200$\\
nozCOSMOS    &$-0.0143 \pm 0.0200$&$-0.0159 \pm 0.0200$&$-0.0106 \pm 0.0200$&$ 0.0053 \pm 0.0200$&$ 0.0135 \pm 0.0200$\\
speczquality4&$-0.0141 \pm 0.0200$&$-0.0163 \pm 0.0200$&$-0.0121 \pm 0.0200$&$ 0.0043 \pm 0.0200$&$ 0.0125 \pm 0.0200$\\
multispec3   &$-0.0158 \pm 0.0200$&$-0.0203 \pm 0.0200$&$-0.0173 \pm 0.0200$&$-0.0033 \pm 0.0200$&$-0.0012 \pm 0.0200$\\
\hline
\hline
\end{tabular}
\end{table*}

\section{Gold sample representation statistics}\label{app: neff}
In this work we have tested the sensitivity of our cosmological parameter estimates to differently constructed gold
samples within KV450. Each of these gold samples produces a subset of the available photometric data, and
results in a different set of tomographic redshift distributions. We present these representation statistics and 
corresponding redshift distributions means here in Table \ref{tab: coverage}. 

The combinations of mean redshift and representation statistics tells an interesting story regarding which photometric
data are being removed by each of our gold sample definitions. There is a clear correlation between the removal of
photometric data and a subsequent decrease in the mean redshift of the tomographic bins. The most obvious examples of
this are in the cases of our noDEEP2 and multispec3 samples, where the gold selection removes $30\%$ and $45\%$ of the
fiducial $n_{\rm eff}$ in the fifth tomographic bin, respectively. 
These samples also show the largest redshift distribution shifts within our gold samples: $\Delta \langle z \rangle
\sim 0.05$ in the fifth tomographic bin. The suggests that the gold-sample definition is preferentially
removing truly high-redshift sources from the photometric sample, as expected. This is indicative of the robustness of
the joint redshift distribution estimation and gold selection; unlike the case of the redshift calibration in 
\citetalias{hildebrandt/etal:2020}, each combination of gold-sample and redshift distribution presented here is
compatible, and differences in sample mean redshifts are not indications of bias in the redshift calibration methodology.
This is an important distinction between the different redshift distributions presented here and in
\citetalias{hildebrandt/etal:2020}. 

Finally, we note the impact that the reduced effective number density of each gold sample has on our posterior
constraint of $S_8$. The multispec3 subsample, for example, has roughly $50\%$ of the photometric $n_{\rm eff}$ of the
KV450-DIR sample per tomographic bin, but shows only a $\sim35\%$ larger uncertainty on $S_8$. This is in agreement with
the results of \citetalias{hildebrandt/etal:2020}, who found that the KV450-DIR $S_8$ uncertainty was limited equally by
statistical and systematic uncertainties. 

\begin{table*}
\centering
\caption{Mean tomographic redshifts and representation statistics of photometric source galaxies within each of our gold
samples. Representation is defined using the effective number density of sources for cosmic shear studies, $n_{\rm
eff}$, in each of the gold samples relative to a reference sample $n_{\rm eff}$.  For the fiducial representation
statistic we use the full KV450 photometric data set for reference (i.e. $\rm n^{\rm fid}_{\rm eff}/n^{\rm all}_{\rm
eff}$), while all other gold sample representations use the fiducial for reference (i.e. $\rm n^{\rm gold}_{\rm
eff}/n^{\rm fid}_{\rm eff}$). The statistics are all given per tomographic bin. The table demonstrates that each of our
nozCOSMOS, noVVDS, and noDEEP2 gold samples has preferentially removed a different section of the colour-space.  This is
joined, however, by a shift in the mean redshift of the tomographic bin, indicating that the loss of the colour redshift
space has been accounted for in the reconstruction.  As expected, the multispec3 selection is highly restrictive,
removing $30-45\%$ of the fiducial photometric $n_{\rm eff}$ in every bin.
}\label{tab: coverage}
\begin{tabular}{c|ccccc|ccccc}
\hline
Gold         & \multicolumn{5}{c|}{$\rm n^{\rm gold}_{\rm eff}/n^{\rm ref}_{\rm eff} (\%)$} &
               \multicolumn{5}{c}{$\langle z \rangle$} \\
Sample       & bin1                    & bin2        & bin3        & bin4       & bin5 & 
               bin1                    & bin2        & bin3        & bin4       & bin5 \\
\hline
KV450-DIR    &  $100.0 $ &  $100.0 $ &  $100.0 $ &  $100.0 $ &  $100.0 $ &
                $0.369 $ &  $0.463 $ &  $0.643 $ &  $0.806 $ &  $0.973 $ \\
Fiducial     &  $ 78.6 $ &  $ 82.1 $ &  $ 79.2 $ &  $ 82.3 $ &  $ 91.6 $ &
                $0.236 $ &  $0.379 $ &  $0.537 $ &  $0.766 $ &  $0.948 $ \\
\hline 
nozCOSMOS    &  $  93.4$ & $  92.2$ & $  92.0$ & $  88.3$ & $  91.5$ &
                $ 0.214$ & $ 0.371$ & $ 0.529$ & $ 0.755$ & $ 0.945$  \\
noDEEP2      &  $  97.7$ & $  96.2$ & $  88.6$ & $  79.5$ & $  72.7$ &
                $ 0.237$ & $ 0.374$ & $ 0.516$ & $ 0.737$ & $ 0.908$  \\
noVVDS       &  $  97.1$ & $  92.4$ & $  86.2$ & $  88.1$ & $  91.4$ &
                $ 0.237$ & $ 0.373$ & $ 0.537$ & $ 0.766$ & $ 0.951$  \\
speczquality4&  $  95.0$ & $  92.0$ & $  87.2$ & $  86.8$ & $  89.4$ &
                $ 0.231$ & $ 0.367$ & $ 0.524$ & $ 0.756$ & $ 0.941$  \\
multispec3   &  $  71.2$ & $  72.7$ & $  65.0$ & $  55.4$ & $  54.5$ &
                $ 0.226$ & $ 0.369$ & $ 0.515$ & $ 0.737$ & $ 0.906$  \\
\hline
\hline
\end{tabular}
\end{table*}

\section{Additional marginal constraints}\label{app: marginals}
\begin{figure*}
\centering
\includegraphics[scale=1.1]{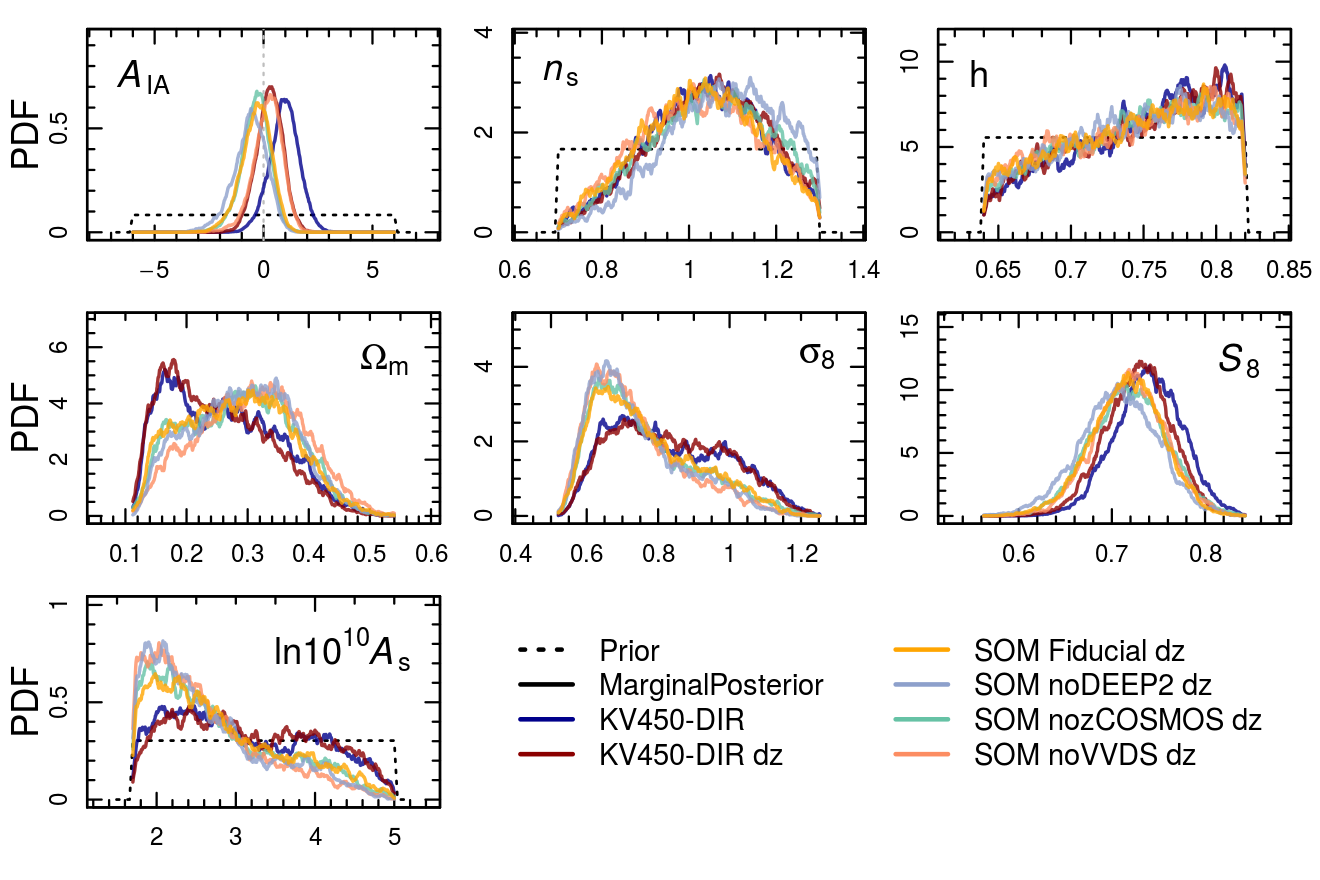}
\caption{Marginal posterior distributions for a subset of the cosmological, nuisance, and derived parameters used in
our cosmological model. Coloured lines represent the marginal distributions from various samples. Dashed lines show 
the priors for all non-derived parameters. There is a clear difference between the marginal distributions of the gold
and full-sample (`KV450-DIR') analyses. We note in particular that the previously observed preference within KV450 
for small values of the matter density parameter $\Omega_{\rm m}$ is removed in our gold analyses. The gold analyses
also prefer a lower value of $A_{\rm IA}$, consistent with $0$ in all cases. }\label{fig: marginals}
\end{figure*}

\begin{figure*}
\centering
\includegraphics[width=\textwidth]{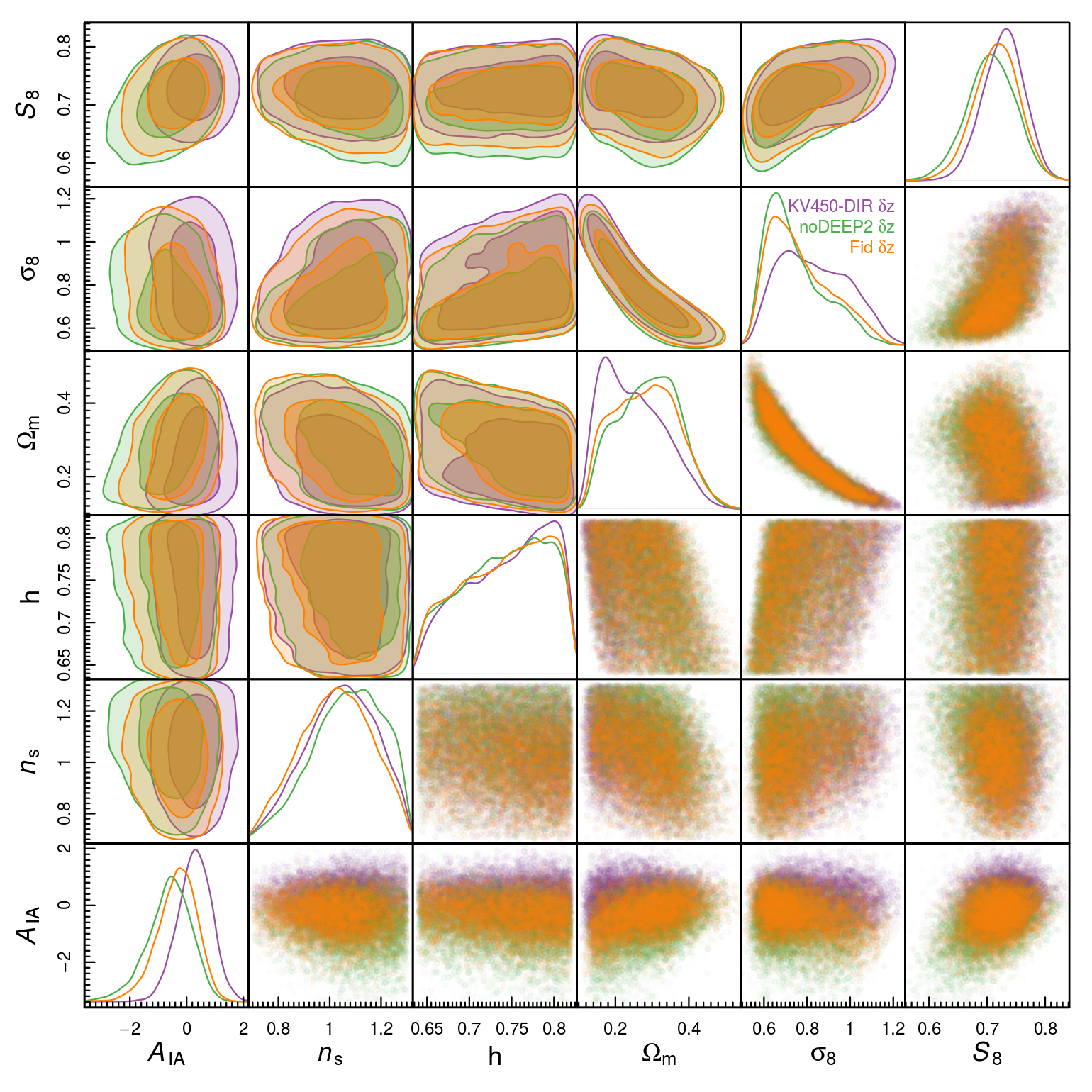}
\caption{\bbf{Marginal and 2D posterior distributions for a subset of the cosmological, nuisance, and derived parameters used in
our cosmological model. We restrict this figure to only the `KV450-DIR $\delta z$' (purple), `SOM-Gold Fidicial $\delta
z$' (gold), and `SOM-Gold noDEEP2 $\delta z$' (green) analyses, as these three analyses encompase the range of
posteriors in all of our gold sample analyses (and show similar degeneracies). The joint posterior of $A_{\rm IA}$ 
and $S_8$ demonstrates that, were a stronger prior on $A_{\rm IA}$ justifiable, agreement between the various gold
samples would be even higher.}}\label{fig: posteriors}
\end{figure*}

Here we present a subset of the posterior constraints from a subset of our gold sample analyses. 
In \bbf{Figures \ref{fig: marginals} and \ref{fig: posteriors}} we show four of the 14 cosmological and nuisance parameters which are used by our
likelihood model ($A_{\rm IA}$, $n_{\rm s}$, $h$, and ${\rm ln}10^{10}A_{\rm s}$), as well as three derived
parameters ($\Omega_{\rm m}$, $\sigma_8$, and $S_8$), \bbf{in their marginal and 2D projections respectively. The marginal}
mean and standard deviations of these posterior distributions are also provided in Table \ref{tab: marginals}. For an in
depth description of the likelihood used here see
\citetalias{hildebrandt/etal:2020}. We have selected these parameters to show as they are of cosmological interest
and/or are not prior dominated in our analysis. \bbf{In particular, we opt not to show the various redshift distribution
bias posteriors, as they follow their priors exactly in all our gold sample analyses. In this way, our results reinforce
those of previous works: cosmic shear correlation functions cannot self-calibrate redshift distribution biases.} 

\begin{table*}
\centering
\caption{Marginal parameter means and standard deviations for the subset of parameters shown in Figure \ref{fig:
marginals}.}\label{tab: marginals}
\begin{tabular}{c|rrrrr|r}
\hline
Parameter &
\multicolumn{1}{c}{KV450-DIR}  & \multicolumn{1}{c}{Fiducial}  & \multicolumn{1}{c}{nozCOSMOS}  &
\multicolumn{1}{c}{noVVDS  }  & \multicolumn{1}{c}{noDEEP2  }  & \multicolumn{1}{c}{KV450-DIR}  \\
& \multicolumn{5}{c}{$\delta z \neq 0$} & \multicolumn{1}{c}{ $\delta z = 0$} \\
\hline
{$A_{\rm IA}$}                  & $0.282\pm0.594$  &$-0.344\pm0.695$  &$-0.366\pm0.650$  & $0.198\pm0.665$  &$-0.627\pm0.775$  & $0.959\pm0.671$ \\
{$ n_{\rm s} $}                 & $1.044\pm0.130$  & $1.023\pm0.133$  & $1.042\pm0.136$  & $1.020\pm0.137$  & $1.072\pm0.130$  & $1.032\pm0.131$ \\
{$ h $}                         & $0.747\pm0.049$  & $0.742\pm0.049$  & $0.743\pm0.049$  & $0.741\pm0.050$  & $0.741\pm0.049$  & $0.748\pm0.048$ \\
{$ {\rm ln}10^{10}A_{\rm s}$}   & $3.158\pm0.864$  & $2.816\pm0.806$  & $2.762\pm0.795$  & $2.617\pm0.715$  & $2.653\pm0.743$  & $3.099\pm0.882$ \\
{$ \Omega_{\rm m} $}            & $0.249\pm0.082$  & $0.282\pm0.085$  & $0.286\pm0.085$  & $0.305\pm0.085$  & $0.291\pm0.081$  & $0.259\pm0.087$ \\
{$ \sigma_8$}                   & $0.834\pm0.156$  & $0.768\pm0.143$  & $0.765\pm0.145$  & $0.739\pm0.132$  & $0.743\pm0.134$  & $0.833\pm0.160$ \\
{$ S_8 $}                       & $0.728\pm0.035$  & $0.716\pm0.038$  & $0.718\pm0.041$  & $0.719\pm0.038$  & $0.707\pm0.042$  & $0.739\pm0.036$ \\
 \hline
 \hline
\end{tabular}
\end{table*}

The marginal distributions from each of our gold samples in Figure \ref{fig: marginals} are in good agreement. Comparing
the various gold sample analyses to our two `KV450-DIR' runs, which use data vectors and redshift distributions
equivalent to those in \citetalias{hildebrandt/etal:2020}, we see some interesting differences. Firstly, we note that
the gold samples no longer demonstrate a preference for small values of the matter density parameter, $\Omega_{\rm
m}\sim0.18$.  Instead, our gold marginal distributions all peak at values $\Omega_{\rm m} \sim 0.3$, in much better
agreement with concordance cosmological parameters. This has a subsequent effect on the marginal distribution of
$\sigma_8$, causing it to be considerably narrower for our gold analysis than in the KV450-DIR cases; we find
$\sigma_8=0.762_{-0.180}^{+0.070}$ compared to $\sigma_8=0.836_{-0.218}^{+0.132}$.  

\bbf{Focussing on the} marginal constraints on $A_{\rm IA}$, we see that this parameter shows the greatest variation
within our gold sample analyses. Interestingly, though, we note that only the results of `KV450-DIR' (i.e. KV450-DIR
using the fiducial redshift bias priors) demonstrate a preference for non-zero values of $A_{\rm IA}$. In all other
cases, the marginal constraints are consistent with $A_{\rm IA}=0$; in agreement with recent work on
intrinsic alignments within KiDS \citep{fortuna/etal:2020}, who advocate $0\leq A_{\rm IA}\leq 0.2$.   

\bbf{Furthermore, we note a correlation between the values of $A_{\rm IA}$ and $S_8$ in the marginal distributions.
Looking to the 2D posteriors in Figure \ref{fig: posteriors}, the degeneracy between these two parameters is clearer.
Interestingly, the $A_{\rm IA}$ versus $S_8$ joint posterior (for the `KV450-DIR', `SOM-Gold Fiducial', and `SOM-Gold
noDEEP2' samples shown) demonstrates that a stronger prior on the value of $A_{\rm IA}$ would likely bring the value of
$S_8$ in these analyses into even closer agreement.  This in turn suggests that the freedom of our intrinsic alignment
parameter does not improve the agreement of $S_8$ between our different gold samples; rather the opposite. For
demonstrative purposes only, we present in Table \ref{tab: S8AIA} the marginal posterior constraints on $S_8$ for
posterior samples within $0\leq A_{\rm IA}\leq 0.2$, compared to the full posterior results (which invokes an intrinsic
alignment amplitude prior of $-6\leq A_{\rm IA}\leq 6$). This clearly demonstrates the interplay between $A_{\rm IA}$
and $S_8$; for the restricted prior, we find a maximal deviation between marginal constraints of $|\Delta S_8|\lesssim
0.005$, only marginally above MCMC noise. Nonetheless, we caution against the over-interpretation of these constraints,
as a stronger prior on $A_{\rm IA}$ is not yet justifiable: analyses such as \citet{fortuna/etal:2020} invoke a number
of assumptions that prohibit the use of their constraints as prior distributions.}

\begin{table}
\caption{\bbf{ For demonstrative purposes only, the marginal constraints of $S_8$ with our fiducial and reduced $A_{\rm IA}$
prior. This demonstrates that, at fixed values of $A_{\rm IA}$, each of the gold samples are in agreement with respect
to $S_8$, and therefore that the differences seen in $S_8$ are not improved by the freedom of $A_{\rm IA}$ in our
cosmological model.}}\label{tab: S8AIA}
\begin{tabular}{cc|rr}
\multicolumn{2}{c}{Gold}   & \multicolumn{2}{c}{Marginal $S_8$ Constraint} \\
\multicolumn{2}{c}{Sample} & $-6\leq A_{\rm IA}\leq 6$ & $0\leq A_{\rm IA}\leq 0.2$ \\
\hline
                  & \multicolumn{1}{c}{KV450-DIR}     & $0.728\pm0.035$  & $0.729\pm0.035$  \\
                  & \multicolumn{1}{c}{Fiducial}      & $0.716\pm0.038$  & $0.724\pm0.036$  \\
$\delta z\neq 0$  & \multicolumn{1}{c}{nozCOSMOS}     & $0.718\pm0.041$  & $0.728\pm0.038$  \\
                  & \multicolumn{1}{c}{noVVDS  }      & $0.719\pm0.038$  & $0.722\pm0.037$  \\
                  & \multicolumn{1}{c}{noDEEP2  }     & $0.707\pm0.042$  & $0.724\pm0.038$  \\
\hline
                  & \multicolumn{1}{c}{KV450-DIR}     & $0.739\pm0.036$  & $0.730\pm0.034$  \\
$\delta z= 0$     & \multicolumn{1}{c}{Fiducial}      & $0.714\pm0.039$  & $0.722\pm0.036$  \\
                  & \multicolumn{1}{c}{speczquality4} & $0.727\pm0.039$  & $0.729\pm0.037$  \\
                  & \multicolumn{1}{c}{multispec3}    & $0.721\pm0.044$  & $0.725\pm0.041$  \\
\hline
\hline
\end{tabular}
\end{table}


%
%
%

\end{document}